\documentclass[11pt]{article}
\usepackage{isolatin1,latexsym,epsfig,a4wide}

\title{The Fundamental Diagram of Pedestrian Movement Revisited}

\author{Armin Seyfried$^1$, Bernhard Steffen$^1$, Wolfram Klingsch$^2$ and Maik Boltes$^1$\\ [0.2cm] 
\footnotesize $^1$Central Institute for Applied Mathematics, Research Centre J\"ulich, 52425 J\"ulich, Germany\\
[-0.1cm]\footnotesize $^2$Institute for Building Material Technology and Fire Safety Science, University of Wuppertal\\ 
[-0.15cm]\footnotesize Pauluskirchstrasse 7, 42285 Wuppertal, Germany\\
\footnotesize  E-mail: A.Seyfried@fz-juelich.de, B.Steffen@fz-juelich.de, klingsch@uni-wuppertal.de, M.Boltes@fz-juelich.de\\
}

\date{\footnotesize First subm. June 27, 2005; accepted September 12, 2005}

\begin{document}

\maketitle

\begin{abstract}
\par 
\noindent
The empirical relation between density and velocity of pedestrian movement 
is not completely analyzed, particularly with regard to the `microscopic' 
causes which determine the relation at medium and high densities. The simplest 
system for the investigation of this dependency is the normal movement of 
pedestrians along a line (single-file movement). This article presents 
experimental results for this system under laboratory conditions and discusses 
the following observations: The data show a linear relation between the velocity 
and the inverse of the density, which can be regarded as the required length of 
one pedestrian to move. Furthermore we compare the results for the single-file 
movement with literature data for the movement in a plane. This comparison shows 
an unexpected conformance between the fundamental diagrams, indicating that 
lateral interference has negligible influence on the velocity-density relation 
at the density domain $1\;m^{-2}<\rho<5\;m^{-2}$. 
In addition we test a procedure for automatic recording of pedestrian flow 
characteristics. We present preliminary results on measurement range and 
accuracy of this method. 
\end{abstract}


\section{Introduction}

Pedestrian dynamics has a multitude of practical applications, like the evaluation 
of escape routes or the optimization of pedestrian facilities, along with some more 
theoretical questions \cite{PED01,PED03,HELB05,OED63,FRUIN71,PRED71}. Empirical studies 
of pedestrian streams can be traced back to the year 1937 \cite{PRED71}. To this 
day a central problem is the relation between density and flow or velocity. This 
dependency is termed the fundamental diagram and has been the subject of many 
investigations from the very beginning
\cite{OED63,PRED71,TOG55,HANK58,OLD68,NAV69,CAR70,WES71,OFL72,POL83,LAM95,HOS04}. 
This relation quantifies the capacity of pedestrian facilities and thus allows 
e.g. the rating of escape routes. Furthermore, the fundamental diagram 
is used for the evaluation of models for pedestrian movement 
\cite{HELB95,SCHA01,SCHR02,HOOG00}, and is a primary test whether the model is 
suitable for the description of pedestrians streams 
\cite{HOOG02b,SCHR02b,SCHA04,RIMEA}.

The velocity-density relation differs for various facilities like stairs, 
ramps, bottlenecks or halls. Moreover one has to distinguish between uni- or 
bi-directional streams. The simplest system in this enumeration is the uni-directional 
movement of pedestrians in a plane without bottlenecks. In this context the fundamental 
diagram of Weidmann \cite{WEID93} is frequently cited. It is a part of a review work 
and the author summarized 25 different investigations for the determination of the 
fundamental diagram. Apart from the fact, that with growing density the velocity 
decreases, the relation shows a non-trivial form. Weidmann notes that different 
authors choose different approaches to fit their data, indicating that the dependency 
is not completely analyzed. A multitude of possible effects can be considered which 
may influence the dependency. For instance we refer to passing maneuvers, internal 
friction, self-organization phenomena or ordering phenomena like the `zipper' 
effect \cite{HOOG05}. A reduction of the degrees of freedom helps to restrict possible 
effects and allows an improved insight to the problem. To exclude in a natural way 
the influences of passing maneuvers, internal friction or ordering phenomena, we 
choose a one-dimensional system. At this we restrict the investigation on the 
velocity-density relation for the normal and not for the pushy or panic movement 
of pedestrians. Similar experiments with the focus on the time gap distribution and 
results for the clogging in pushy crowds can be found in \cite{HELB05}.

Furthermore, we test a procedure for automatic recording of pedestrian flow 
characteristics. This method uses stereo video processing and allows the determination 
of trajectories of individual persons. We present preliminary results on measurement 
range and accuracy of this method.


\section{Movement in a plane}

In the field of pedestrian dynamics there are two different ways to quantify the 
capacity of pedestrian facilities. Either the relation between density $\rho$ and 
velocity $v$ or the relation between density and flow $\Phi = \rho\,v$ is 
represented. Following Weidmann we choose the presentation of density and velocity. 

\begin{figure}[thb]
  \epsfxsize=11.5cm
  \centerline{\epsfbox{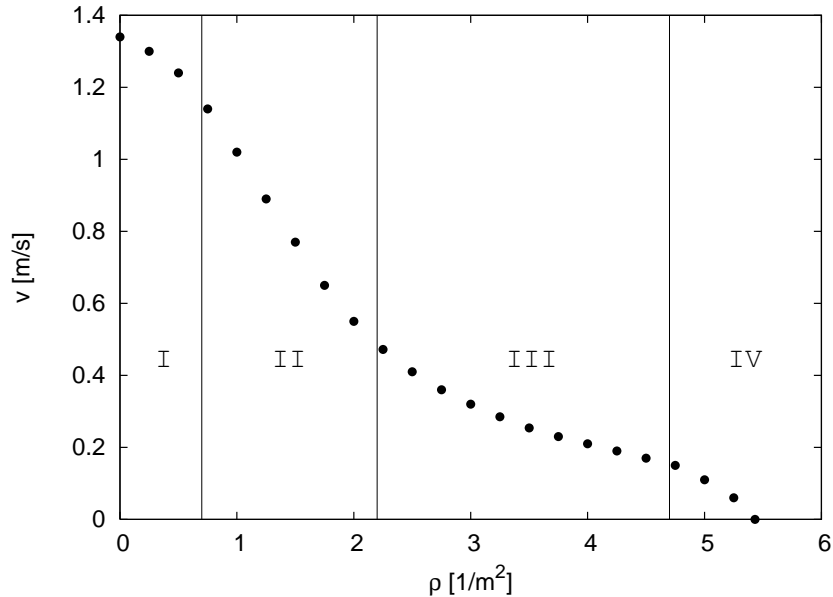}}
  \caption{Empirical relation between density and velocity according to Weidmann 
(Page 52 in \cite{WEID93}). The partition refers to domains with qualitative 
different decrease of the velocity.\label{WEID}}
\end{figure}

Figure \ref{WEID} shows the empirical velocity-density relation for pedestrian 
movement in a plane according to Weidmann\footnote{Note 
that Weidmann's combination does not distinguish between uni- (e.g. \cite{HANK58}) 
and bi-directional (e.g. \cite{OLD68}) movement}. The maximal density in this diagram 
is $5.4/m^2$, though some authors \cite{PRED71,TOG55,WES71} have found higher densities. 
The slope varies for different density-domains indicating diverse effects which reduce 
the velocity. We discuss possible causes for this slope-variation by means of the 
Level of Service concept (LOS) \cite{OED63,FRUIN71,WEID93}.

\begin{description}
\item[Domain I $\rho<0.7$] At low densities there is a small and increasing decline 
of the velocity. The velocity is mostly determined by the individual free velocity 
of the pedestrians. Passing maneuvers for keeping the desired velocity are possible. 
The decrease is caused by the passing maneuvers. 
\item[Domain II $0.7\leq\rho<2.3$] The velocity decrease is nearly linear with growing 
density. Due to the reduction of the available space, passing maneuvers of slower 
pedestrians are hardly feasible and the possibility to choose the desired velocity 
is restricted. At least at uni-directional streams the available space is large 
enough to avoid contacts with other pedestrians. Thus the internal friction can only 
be of negligible influence.
\item[Domain III $2.3\leq\rho<4.7$] The linear decrease of the velocity ends and the 
curvature changes. For growing density the velocity remains nearly constant. Contacts 
with other pedestrians are hardly avoidable. While the internal friction increases 
compared with domain II, the reduction of the velocity diminishes. 
\item[Domain IV $\rho\geq4.7$] The velocity declines rapidly. The available space is 
strongly restricted. Internal friction may be a determining factor. There has to be 
a maximal density because the pedestrians behave like hard bodies. 
\end{description}

As pointed out, there are several hints to effects which can influence the 
reduction of the velocity at different densities. But it is not clear which 
`microscopic' properties of pedestrian movement lead to the nearly linear decrease of 
the velocity and to a slower decrease at high densities. Particularly the possible 
influence of the internal friction is contradictory to the slope of the 
velocity-density relation. We have to clarify the influences of collective 
behavior like marching in lock-step \cite{NAV69} or self-organization phenomena 
like the 'zipper' effect, which can be observed at bottlenecks \cite{HOOG05}. 


\section{Single-file movement}

\subsection{Description of the experiment}

Our target is the measurement of the relation between density and velocity for the 
single-file movement of pedestrians. To facilitate this with a limited amount of test 
persons also for high densities and without boundary effects, we choose a 
experimental set-up similar to the set-up in \cite{HANK58}.

\begin{figure}[thb]
  \epsfxsize=11cm
  \centerline{\epsfbox{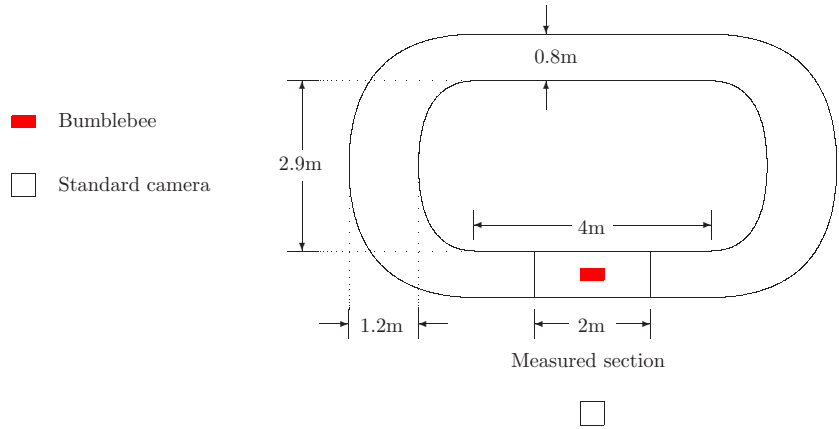}}
  \caption{Experimental set-up for the measurement of the velocity-density 
relation for the single-file movement. \label{SKIZ}}
  \end{figure}

The corridor, see Figure \ref{SKIZ}, is build up with chairs and ropes. The width of 
the passageway in the measurement section is $0.8\,m$. Thus passing is prevented and 
the single-file movement is enforced. The width of the corridor is not important as 
long it does not impede the free movement of the arms and on the other hand does not 
allow the formation of two - possible interleaving - lanes.
The circular guiding of the passageway gives 
periodic boundary conditions. To reduce the effects of the curves on the measurement, 
we broaden the corridor in the curve and choose the position of the measured section 
in the center of the straight part of the passageway. The length of the measured 
section is $l_m=2\,m$ and the whole corridor is $l_p=17.3\,m$. The experiment is 
located in the auditorium Rotunde at the Central Institute for Applied Mathematics 
(ZAM) of the Research Centre J\"ulich. The group of test persons is composed of 
students of Technomathematics and staff of ZAM. To get a normal and not a pushy movement 
the test persons are instructed not to hurry and omit passing. 
This results in a rather relaxed conduct, i. e. the 
resulting free velocities are rather low. Most real life situations will 
show higher free velocities and more pushing. To enable measurements at different 
densities we execute six cycles with $N=1,15,20,25,30,34$ numbers of test persons 
in the passageway. For the cycle with $N=1$ every person passes alone through the 
corridor. For the other cycles we first distribute the persons uniformly in the 
corridor. After the instruction to get going, every person passes the passageway two 
to three times. After that we open the passageway and let the test persons get out. 
We observed that the test persons walked straight through the measurement section
and follow each other up to a few centimeter of lateral variation. This does not mean 
that they walked exactly on the same path but lateral fluctuations are highly correlated.
This observation is reinforced by the documentation of marching in lock-steps, which will
be discussed later.

\subsection{Measurement set-up}

\begin{figure}[thb]
  \epsfxsize=9cm
  \centerline{\epsfbox{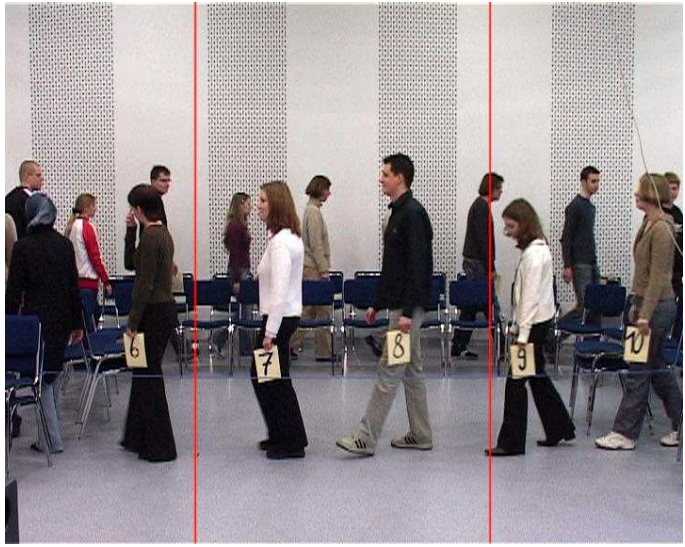}}
  \caption{One frame of the standard video recording of the cycle with $N=20$ after
the preparation for the manual analysis. \label{PHOTO}}
\end{figure}

For the measurement of the flow characteristics we use both a manual and an automatic 
procedure. The manual procedure is based on standard video recordings with a DV camera 
(PAL format, 25 fps) of the measured section. These recordings are prepared on a 
computer to show time, frame number and the limits of the measured section. 
After the preparation the recordings were analyzed frame-wise, thus the accuracy of 
the extracted time is $0.04\,s$. To minimize the errors for extracting time data, 
the collected time is transferred by cut and paste between the video editing system 
and the spread sheet. Figure \ref{PHOTO} shows one frame of the cycle with $N=20$ 
persons. For every test person $i$ we collect the entrance time (of the ear) in the 
measurement section $t_i^{in}$ and the exit time $t_i^{out}$. To ease the assignment 
of times the test persons carry numbers.

Additionally we test an automated procedure, based on a commercial system of 
Point Grey Research \cite{PGR}. The system is composed of the Bumblebee stereo vision 
camera and the software packages Digiclops, Triclops and Censys3d. The software uses 
stereo recordings for detection and tracking of peoples. The resulting trajectories 
allow the analysis of pedestrian movement in space and time. For this measurement 
we use a bumblebee BW-BB-20. The analysis device is an IBM ThinkPad T30 with 
an Intel Pentium 4 Mobile CPU (512 MB RAM) and the operating system WIN XP SP1. 
The data transfer from the camera to the analysis device is realized via FireWire and a 
PCMCIA FireWire card. Following the recommendation of the manufactures we decide to 
process the data directly without storing the pictures on a hard disk. In our 
measurement setup we are bounded on a transfer rate to the analysis device of about
2 times 20 fps with a resolution of 320 x 240 pixels. 

\subsection{Data analysis}

The manual analysis uses the entrance and exit times $t_i^{in}$ and $t_i^{out}$. These 
two times allow the calculation of the individual velocities 
$v^{man}_i=l_m/(t_i^{out}-t_i^{in})$ and the momentary number $n(t)$ of persons 
at time $t$ in the measured section. The concept of a momentary `density' in the 
measurement region is problematic because of the small (1-5) number of persons involved. 
$\tilde{\rho}^{\,man}(t)= n(t)/l_m$ jumps between discrete values. For a better 
definition we choose  $\rho^{man}(t)=\sum_{i=1}^{N} \Theta_i(t)/l_m$, where $\Theta_i(t)$ 
gives the `fraction' to which the space between person $i$ and 
person $i+1$ is inside. 

\begin{equation}
 \Theta_i(t) =  \left\{\begin{array}{r@{\quad:\quad}l}
\frac{t-t_{i}^{in}}{t_{i+1}^{in}-t_{i}^{in}} & t \in [t_{i}^{in},t_{i+1}^{in}]\\
1 & t \in [t_{i+1}^{in},t_i^{out}] \\
\frac{t_{i+1}^{out}-t}{t_{i+1}^{out}-t_{i}^{out}} &  t \in [t_{i}^{out},t_{i+1}^{out}]\\
0 & \mbox{otherwise}
\end{array} \right. 
\label{DENSDEF} 
\end{equation}

In Figure \ref{PHOTO} the `fractions' are $\Theta_6(t) \approx 0.6$, $\Theta_7(t) =1$, 
$\Theta_8(t) \approx 0.8$, $\Theta_9(t)=0$, resulting in  $\rho(t) \approx 1.2\,m^{-1}$. 
This is 3 persons per $2.5\,m$, which is about the distance between person number 6 and 
number 9. The time average of $\tilde{\rho}^{\,man}$ and $\rho^{man}$ is almost the same. 

\begin{figure}[thb]
  \epsfxsize=11.5cm
  \centerline{\epsfbox{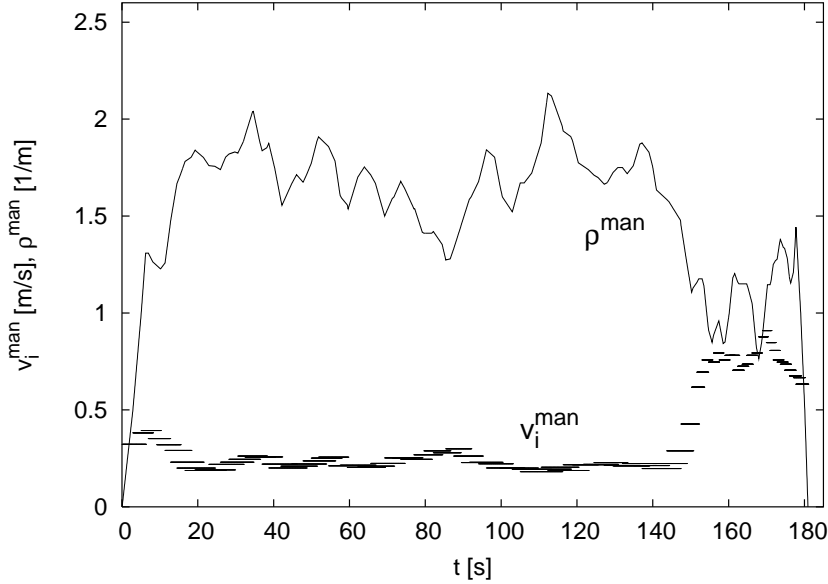}}
  \caption{Time development of the density and individual velocities for the cycle with 
$N=30$. The upper line shows the density,  the lower bars label the time-slots in which 
the pedestrian $i$ is in the measured section and the associated mean velocity $v_i$. 
The fluctuations of the individual velocities are correlated with the density. We see 
the influence of the starting phase and the rearrangement of the boundary conditions due 
to the opening of the passageway at the end of the cycle. \label{t-ri}}
\end{figure}

Figure \ref{t-ri} shows the development of the density and velocity of individual 
persons in time. At the beginning the pedestrians do not react simultaneously and 
they have to tune their movement. After the tuning phase one sees slow 
moderate size fluctuations 
of the density and the velocity. It becomes obvious that these `microscopic' 
fluctuations are correlated. After the opening of the passageway the density declines and 
as a consequence the velocity grows. To consider this correlation in the analysis we 
regard a crossing of an individual pedestrian $i$ with velocity $v^{man}_i$ as one 
statistical event, which is associated to the density $\rho^{man}_i$. While 
$\rho^{man}_i$ is the mean value of the density during the time-slice 
$[t_i^{in},t_i^{out}]$. For this analysis we exclude the part of the data where the 
influence of the tuning phase and the rearrangement of the boundary conditions is 
explicit.

\begin{figure}[thb]
  \epsfxsize=11.5cm
  \centerline{\epsfbox{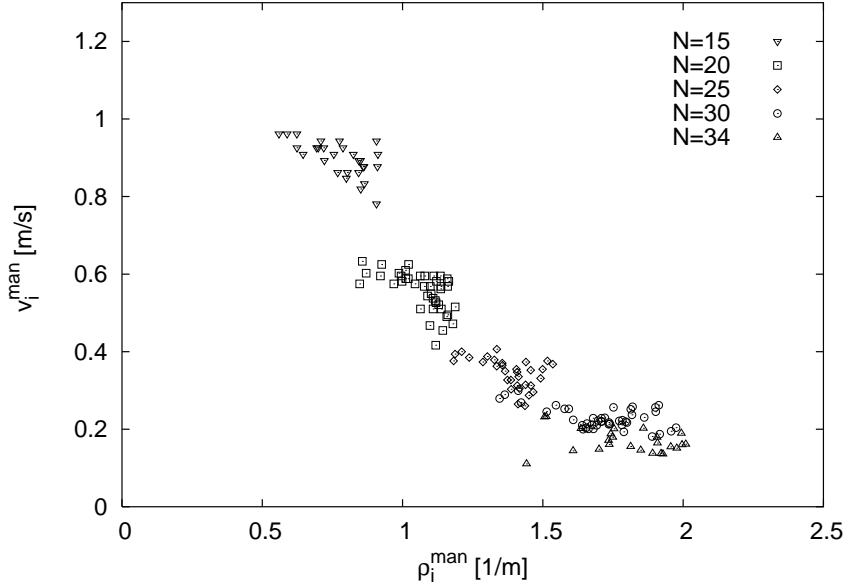}}
  \caption{Dependency between the individual velocity and density for the cycles with 
$N=15,20,25,30$ and $34$.
\label{FD-1D}}
\end{figure}

Figure \ref{FD-1D} shows the distribution of the events $(v^{man}_i,\rho^{man}_i)$ of 
the cycles with $N=15,20,25,30$ and $34$. The cycles with  $N=15,20$ and $25$ result 
in clearly separated areas of densities and velocities. These areas blend for the 
cycles with $N=30$ and $34$. 

\begin{figure}[thb]
  \epsfxsize=11.5cm
  \centerline{\epsfbox{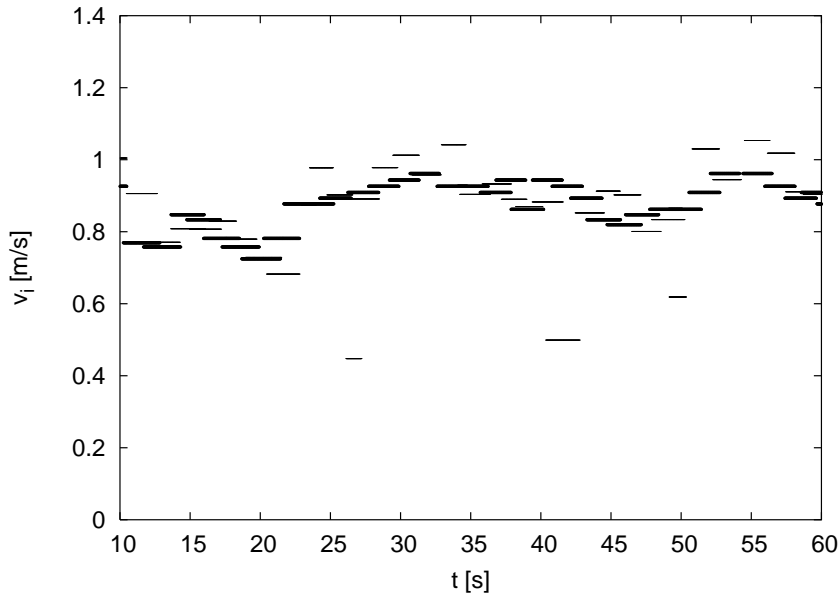}}
  \caption{Comparison for the cycle with $N=15$. The thick lines represent the 
time-slots and velocity determined through the manual procedure. The thin lines 
results from the trajectories gained by the automated procedure. The values for 
the velocities agree roughly, but the results are not comparable on a 
microscopic level.
\label{COMP}}
\end{figure}

The automated procedure uses the trajectories for the analysis. The accuracy of the 
trajectory measurement depends on many factors, like the size of the covered area, 
the capability of the analysis device and the settings of parameters in 
software Censys3d for people detection and tracking. In test-runs we optimize these 
parameters for normal and not too crowded conditions. But during the experiment we 
realized that the system is too sensitive and with growing density the mismeasurements 
increase. For example we observe the loss of trajectories or persons with two 
trajectories, see figure \ref{COMP}. Due to the mismeasurements the determination of densities or other 
microscopic values like the distance between the pedestrians is not reliable. 
The resulting trajectories are thus only appropriate for the calculation of the 
mean value of the velocity. For this purpose we determine for every crossing pedestrian 
$i$ the first $(x_i^{f},t_i^{f})$ and the last point $(x_i^{l},t_i^{l})$ of the 
trajectory in the measured section and calculate the velocity through 
$v^{aut}_i=(x_i^{l}-x_i^{f})/(t_i^{l}-t_i^{f})$. 

In the next step we test if the mismeasurements lead to systematic errors in the 
determination of the velocities and cross-check the data from the manual analysis. 
For this we exclude the data where the influence of the starting phase and 
opening of the passageway are apparent. To reduce the influence of mismeasurements we 
take into account only these trajectories, which last for at least one second.

For the test we determined the following values. Based on the individual velocities 
$v^{aut}_i$ and $v^{man}_i$ we calculate the mean value $v^{man}$ and 
$v^{aut}$ over all individual velocities for the different cycles. Furthermore 
we determine the mean value of the density $\rho^{man}$ during one cycle 
based on the manual analysis and compare them with the densities calculated through 
$\rho = N/l_{p}$, where $l_p$ is the length of the passageway. The values are 
summarized in the following table.

\begin{table}[h]
\begin{center}
\begin{tabular}{| c | c | c | c | c |}
\hline
$N$ &  $\rho [1/m]$ & $\rho^{man} [1/m]$ & $v^{man} [m/s]$ & $v^{aut} [m/s]$ \\ \hline
\hline
1  &      &             & 1.24  (0.15) & 1.37  (0.21) \\ \hline
15 & 0.87 & 0.77 (0.12) & 0.90  (0.05) & 0.88  (0.15) \\ \hline
20 & 1.16 & 1.07 (0.11) & 0.56  (0.05) & 0.51  (0.11) \\ \hline
25 & 1.45 & 1.39 (0.12) & 0.35  (0.04) & 0.26  (0.14) \\ \hline
30 & 1.73 & 1.71 (0.17) & 0.23  (0.03) & 0.16  (0.15) \\ \hline
34 & 1.97 & 1.76 (0.24) & 0.17  (0.03) & 0.13  (0.34) \\ \hline
\end{tabular}
\end{center}
\caption[dummy]{Comparison of mean values and standard deviations ($\sigma$) gained from 
the automated and manual procedure. \label{MEAN}}
\end{table}

Aside from local fluctuations, possible reasons for the deviations of the 
densities $\rho^{man}$ and $\rho$ are the influence of 
the curve and the broadening of the passageway in the curve. The mean velocities 
calculated automatically agree roughly with the velocities extracted 
from the manual procedure. We get a very good agreement for medium densities 
$N = 15,20$. Notable deviations arise at high velocities and high densities. 
The mismeasurements caused by loosing and picking up of trajectories increase 
the fluctuations ($\sigma$). The deviation of the mean value for the cycle with 
$N=1$ from the literature value $v_{free}=1.34\,m/s$ according to Weidmann, can 
be explained by the instruction to the test persons not to hurry.


\section{Results}

In the following we focus on the distance between the pedestrians. For the 
single-file movement the distance to the next close-by pedestrian can be regarded as 
the required length $d$ of one pedestrian to move with velocity $v$. Considering that 
in a one-dimensional system the harmonic average of this quantity is the inverse of 
the density, $d=1/\rho$, one can investigate the relation between required length and 
velocity by means of the velocity-density relation for the single-file movement.

\begin{figure}[thb]
  \epsfxsize=11.5cm
  \centerline{\epsfbox{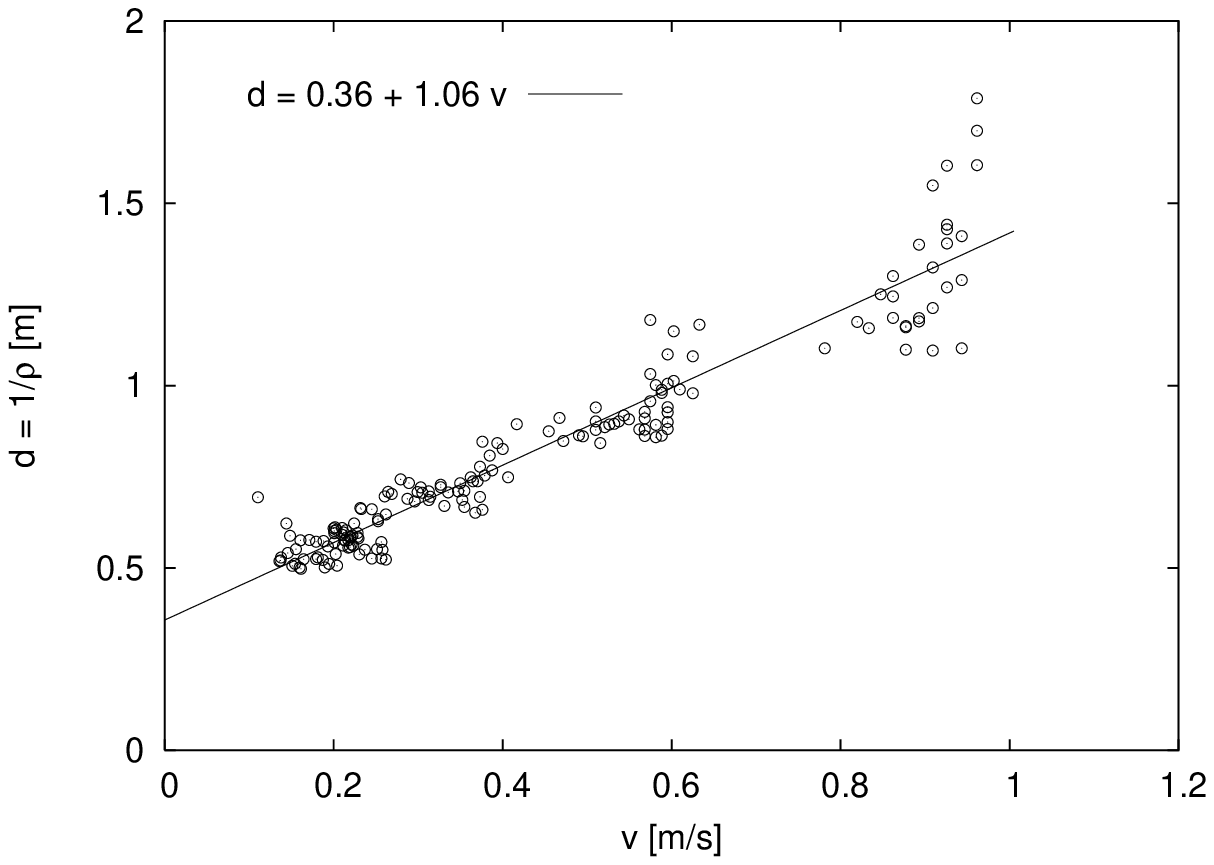}}
  \caption{Dependency between required length and velocity according 
to the data from the cycles with $N=15,20,25,30$ and $34$. We found that a 
linear relationship gives the best fit to the data. \label{LINFIT}}
\end{figure}

Figure \ref{LINFIT} shows the dependency between required length and velocity.
We tested several approaches for the function $d=d(v)$ and found that a 
linear relationship with $d=0.36+1.06 \, v$ gives the best fit to the data. 
According to \cite{WEID93} the step length is a linear function of the velocity\footnote{Lower 
average velocities arise from a lower step frequency.} 
only for  $v \ge \approx 0.5\,m/s$. Thus it is surprising, 
that the linearity for entire distance holds even and persists for velocities 
smaller than $0.5\,m/s$. Possible explanations will be discussed later.

To compare the relation between velocity and density of the single-file movement 
$(1d)$ with the movement in a plane $(2d)$, we have to transform the line-density 
to a area-density. 

\begin{equation}
 \rho_{1d\rightarrow2d} = \rho_{1d}/c(v) \quad \mbox \quad  
 c(v) = \alpha + \beta\;v 
\label{RHOSKAL} 
\end{equation}

The correction term $c(v)$ is introduced to take into account that the lateral 
required width is as well as the longitudinal required length $d$ a function 
of the velocity. For a first approximation we assume a linear dependency. 
According to \cite{WEID93} the mean value of the width of the human body is 
$0.46\;m$ and thus we set $\alpha=0.46\;m$. The parameter $\beta$ considers that with
increasing velocity the lateral required width grows. 

\begin{figure}[thb]
  \epsfxsize=11.5cm 
  \centerline{\epsfbox{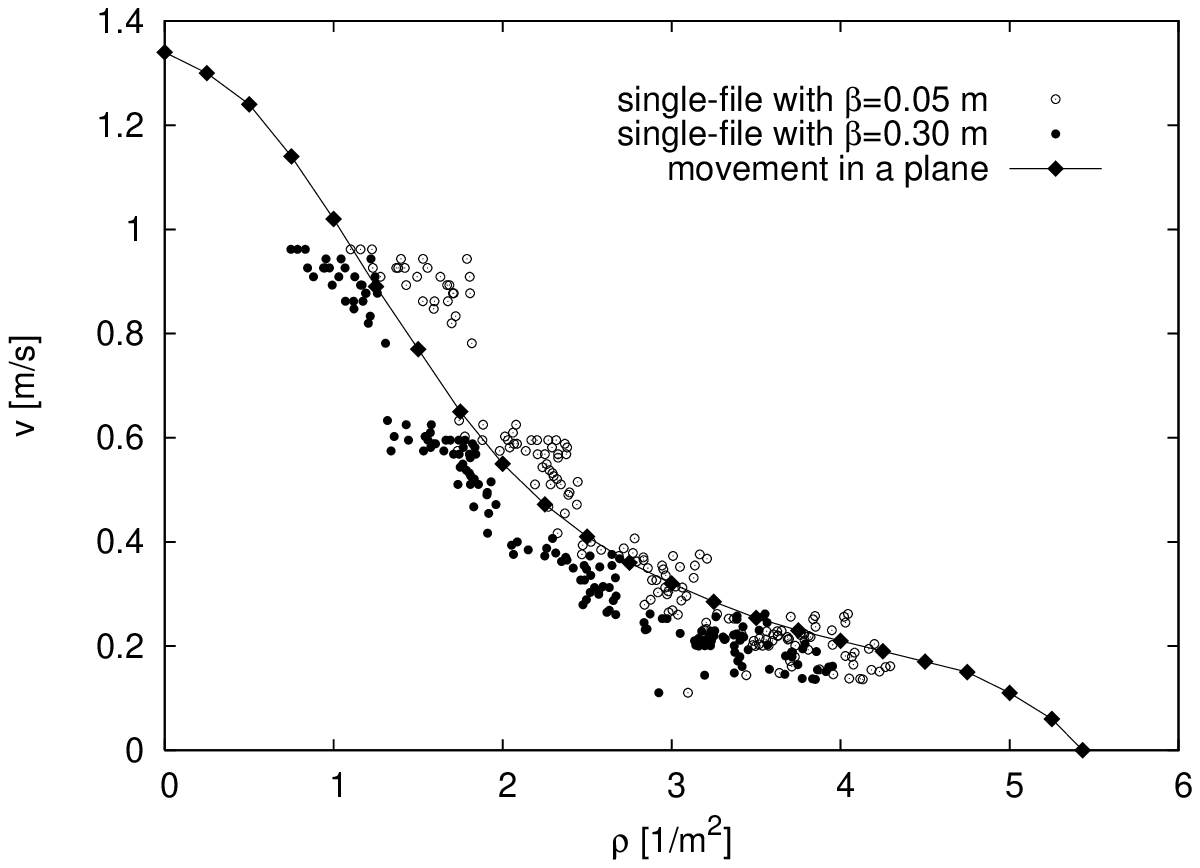}}
  \caption{Comparison of the velocity-density relation for the single-file movement 
with the movement in a plane according to Weidmann \cite{WEID93}. For the scaling of the 
line-density we choose $\beta=0.05\;s$ and $\beta=0.3\;s$, see Equation \ref{RHOSKAL}.
\label{1D-2D}}
\end{figure}

The comparison of the relation between velocity and density for the single-file 
movement with the movement in a plane according to Weidmann shows a surprising 
conformity, see Figure \ref{1D-2D}. So far the value of $\beta$ is unknown, 
but the figure shows, that for a large domain of reasonable values we have an 
obvious accordance between the fundamental diagrams. For the lower bound of 
$\beta=0.05\;s$ one gets a lateral required width at $v=1.3\;m/s$ of $c=0.5\;m$, 
which is to definitely too small. For the upper bound we choose $\beta=0.3\;s$ which 
results in $c(1.3\;m/s)=0.9\;m$ which is certainly too high. But it becomes apparent, 
that the qualitative shape of the velocity-density relations and the magnitude of the 
velocities agree beetween the upper and the lower bound of $\beta$. This conformance 
indicates that two-dimensional specific properties, like internal friction and other 
lateral interferences, have no strong influence on the fundamental diagram at least 
at the density domains considered.

\begin{figure}[thb]
  \epsfxsize=9cm
  \centerline{\epsfbox{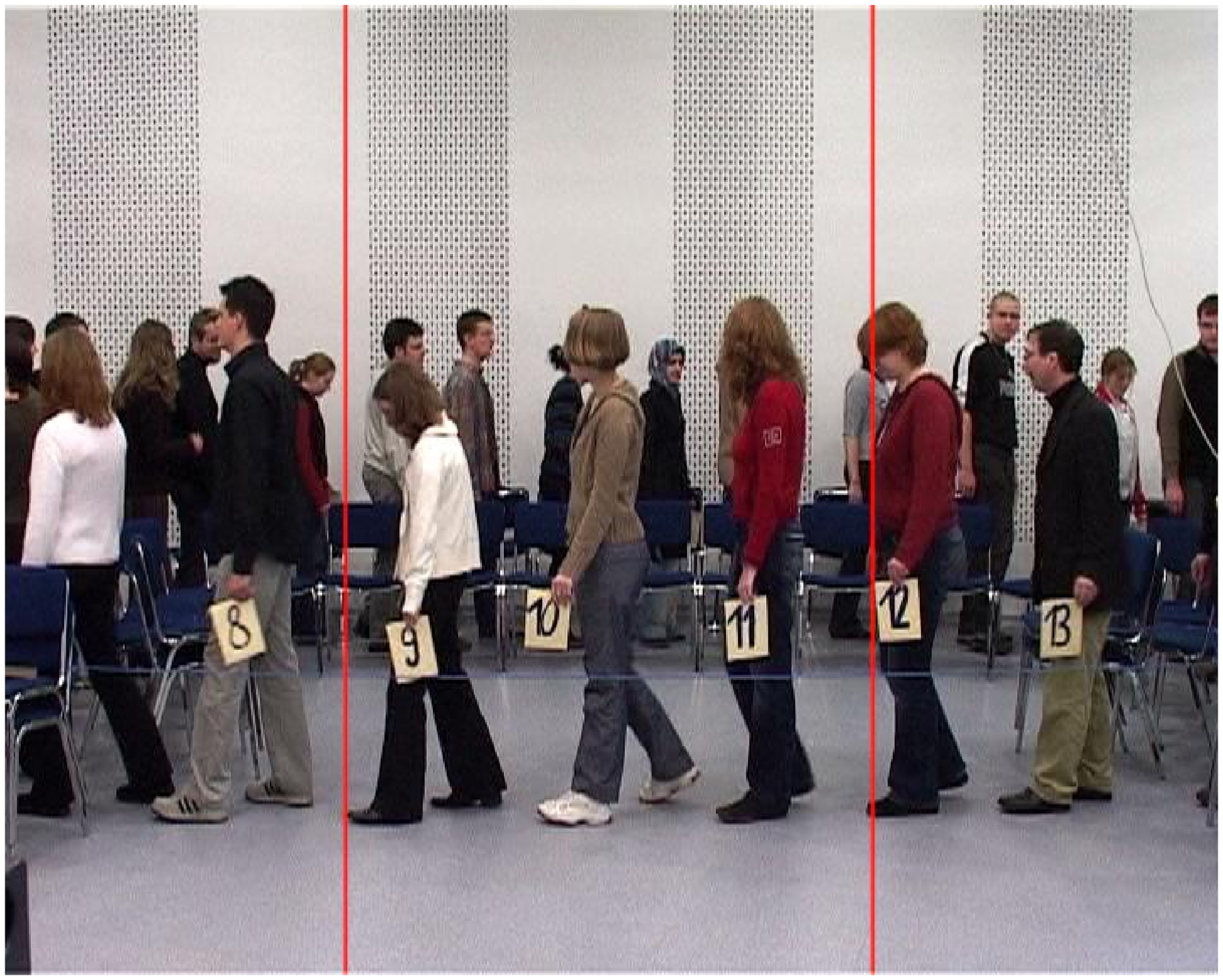}}
  \caption{Example for marching in lock-step for the cycle with $N=30$.\label{PHGL}}
\end{figure}

Instead, the visual analysis of the video recordings suggests that the 
following `microscopic' properties of pedestrian movement determine the relation 
between velocity and density. At intermediate densities and velocities the step 
length is reduced with increasing density. The distance to the pedestrian in front 
is related to the step length as well as to the safety margin to avoid contacts with 
the pedestrian in front. Both, step length and safety margin are connected with the 
velocity. At high densities and small velocities we observed that small groups pass into 
marching in lock-step, see Figure \ref{PHGL}. Furthermore the utilization of the 
available place is optimized. This is achieved by some persons setting their feet 
far right and left of the line of movement, giving some overlap in the space 
occupied with the pedestrian in front. While at intermediate densities and relative 
high velocities the pedestrians are concentrated on their movement, this concentration 
is reduced at smaller velocities and leads to a delayed reaction on the movement of 
the pedestrian in front. The marching in lock-steps and the optimized utilization of 
the available space, which compensate the slower step frequency, are possible 
explanation that the linearity between the required lenght and the velocity holds 
even, see Figure \ref{LINFIT}.

\section{Summary and outlook}

In the investigation presented we determine the fundamental diagram for the 
single-file movement of pedestrians under laboratory conditions. The data are 
appropriate to test, if microscopic models are able to reproduce the empirical 
relation between velocity and density in the simplest system. 

The test of the automated procedure shows that this method is in principle 
capable to measure characteristics of pedestrian movement. The mean values 
of the velocities for different densities acquired automatically are comparable 
with those of the manual data analysis. To facilitate the measurement of 
microscopic characteristics we plan to increase the resolution of the stereo 
recordings and the transfer rate to the analysis system. 

The comparison of the velocity-density relation for the single-file movement with 
the literature-data for the movement in a plane shows a surprising 
agreement for $1\;m^{-2}<\rho<5\;m^{-2}$. The conformance indicates that the internal 
friction and other lateral 
interferences, which are excluded in the single-file movement, have no influence 
on the relation at the density domains considered. The visual analysis of 
the video recording give hints to possible effects, like the self-organization 
through marching in lock-step, the optimized utilization of the available space 
at low velocities and the velocity dependence of step-length and safety margin.
The investigation of the dependency between the required length of one pedestrian 
and velocity indicates a linear relation. For the domain $0.1\,m/s<v<1.0\,m/s$ 
we obtain $d=0.36+1.06 \, v$. The investigation of the interplay between the 
self-organization effects and required length and thus a detailed quantification 
of these effects is necessary.

For real life situation including bottlenecks and changes in direction, and for higher 
densities, however, two-dimensional effects can certainly not be disregarded. Moreover, 
for most laboratory experiments, the emotional situation of the participants is 
relaxed, so some effects of real life will not be shown, as already observed in 
\cite{KOSH83}. Further studies will be required, and some of these will be vastly 
easier after the improvement of the automated procedure.\\ 

\newpage

{\bf Acknowledgments}\\

We thank Patrick Hartzsch for extracting the times from the video, Dr.~Wolfgang Meyer for the advice with statistics, Oliver Passon for careful 
reading and Thomas Lippert for discussions.

\end{document}